\documentclass[aps,preprint,prl]{revtex4}

\usepackage{graphicx}
\usepackage{amsmath}
\usepackage{amssymb}
\usepackage{units}
\usepackage{times}

\begin{document}

%
% Front matter
%

\title{Atomic Hole Doping of Graphene}

\author{Isabella Gierz$^1$}
\author{Christian Riedl$^1$}
\author{Ulrich Starke$^1$}
\author{Christian R.\ Ast$^1$}
\email[Corresponding author; electronic address:\
]{c.ast@fkf.mpg.de}
\author{Klaus Kern$^{1,2}$}
\affiliation{$^1$ Max-Planck-Institut f\"ur Festk\"orperforschung, D-70569 Stuttgart, Germany\\
$^2$ Institut de Physique des Nanostructures, Ecole Polytechnique F{\'e}d{\'e}rale de Lausanne, Ch-1015 Lausanne, Switzerland}

\date{\today}

\maketitle

%
% Body of the article
%

{\bf Graphene is an excellent candidate for the next generation of
electronic materials due to the strict two-dimensionality of its
electronic structure as well as the extremely high carrier
mobility \cite{Novoselov, de Heer, Geim}. A prerequisite for the 
development of graphene based electronics is the reliable control 
of the type and density of the charge carriers by external (gate) 
and internal (doping) means. While gating has been successfully 
demonstrated for graphene flakes \cite{Novoselov, Oostinga, Zhang} 
and epitaxial graphene on silicon carbide \cite{Kedzierski, Gu}, 
the development of reliable chemical doping methods turns out to 
be a real challenge. In particular hole doping is an unsolved issue. 
So far it has only been achieved with reactive molecular adsorbates, 
which are largely incompatible with any device technology. Here we 
show by angle-resolved photoemission spectroscopy that atomic doping 
of an epitaxial graphene layer on a silicon carbide substrate with 
bismuth, antimony or gold presents effective means of p-type doping. 
Not only is the atomic doping the method of choice for the internal 
control of the carrier density. In combination with the intrinsic n-type 
character of epitaxial graphene on SiC, the charge carriers can be tuned 
from electrons to holes, without affecting the conical band structure.}

\begin{figure}
  \includegraphics[width = 0.6\columnwidth]{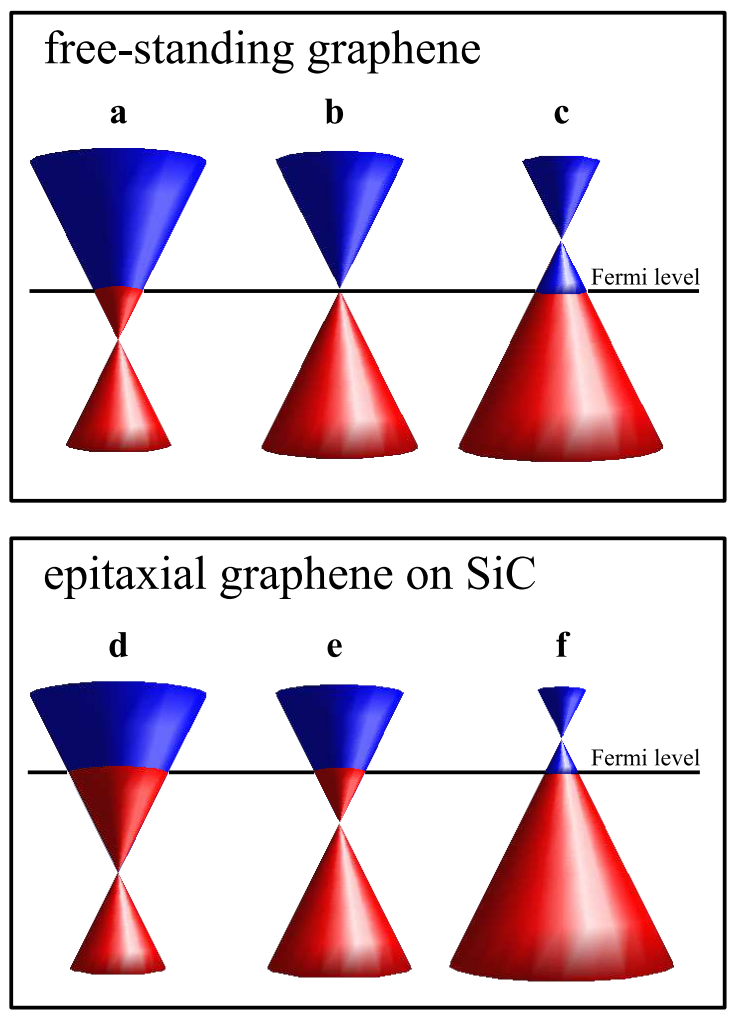}
  \caption{{\bf Doping graphene:} position of the Dirac point and the Fermi level of pristine and epitaxial graphene
  as a function of doping. The upper and lower panels show a free-standing graphene layer and
  an epitaxial graphene layer on silicon carbide, respectively. The left and right panels visualize n-type
  and p-type doping, respectively, while the center panels show the pure graphene layers. For
  the epitaxial graphene a natural substrate induced n-type doping is present.}
  \label{fig:Schematic}
\end{figure}

The recent interest in graphene --- single layers of graphite ---
is based on its peculiar electronic structure. Two-dimensional
by nature, it is a zero-gap semiconductor, i.~e. a semimetal, with
a conically shaped valence and conduction band reminiscent of
relativistic Dirac cones for massless particles
\cite{Wallace,Slonczewski}. As this kind of band structure
provides great potential for electronic devices, one of the key
questions is how to dope the electronic structure with electrons
or holes appropriately for the different devices (see Fig.\
\ref{fig:Schematic}). The semimetallic character induced by the
close proximity of valence and conduction band as well as the
conical shape of the bands result from a delicate balance between
the electrons and the lattice. The challenge here is to interact
with the system just enough to add or remove electrons but not too
much so as to modify or even collapse the electronic structure.
Therefore, it is not an option to replace atoms within the
graphene layer, as is common practice when doping silicon.

For graphene the doping is usually realized by adsorbing atoms and/or
molecules on its surface, i.~e. surface transfer doping
\cite{Maier,Chakrapani,Ristein,Sque}. For n-type (p-type) doping
the electrons have to be easily released into (extracted out of)
the graphene layer. As alkali atoms easily release their valence
electron, they very effectively induce n-type doping \cite{Bostwick,Ohta} (see
Fig.\ \ref{fig:Schematic}(d)). The Dirac point, where the apices
of the two conically shaped bands meet, is shifted further into
the occupied states away from the Fermi level. However, aside from
the fact that epitaxial graphene on silicon carbide is naturally
n-doped, alkali atoms are very reactive and their suitability in
electronic devices is more than questionable.

P-type doping for graphene is quite a bit more challenging. Many
of the elements with a high electronegativity --- e.~g. nitrogen,
oxygen, or fluorine --- form strong dimer bonds. They would not
likely form a stable adlayer on the graphene surface. Therefore,
different molecules such as NO$_2$, H$_2$O, NH$_3$, or the charge
transfer complex tetrafluoro-tetracyanoquinodimethane (F4-TCNQ)
have been used to induce p-type doping in graphene (see Fig.\
\ref{fig:Schematic}(f)) \cite{Chen,Hwang,Wehling,Zhou3}. However,
NO$_2$, H$_2$O and NH$_3$ are very reactive chemicals and therefore not suitable for use in an
electronic material. F4-TCNQ on the other hand plays an important
role in optimizing the performance in organic light emitting
diodes \cite{Blochwitz,Zhou2}, but would also be difficult to implement in large scale fabrication. A viable alternative is
presented by the heavier elements, which are not as reactive as,
e.~g. oxygen or fluorine. Although not obvious, because their
electron affinity is somewhat lower than that of atomic carbon,
bismuth as well as antimony turn out to be able to extract
electrons out of the graphene sheet.

\begin{figure*}
  \includegraphics[width = \columnwidth]{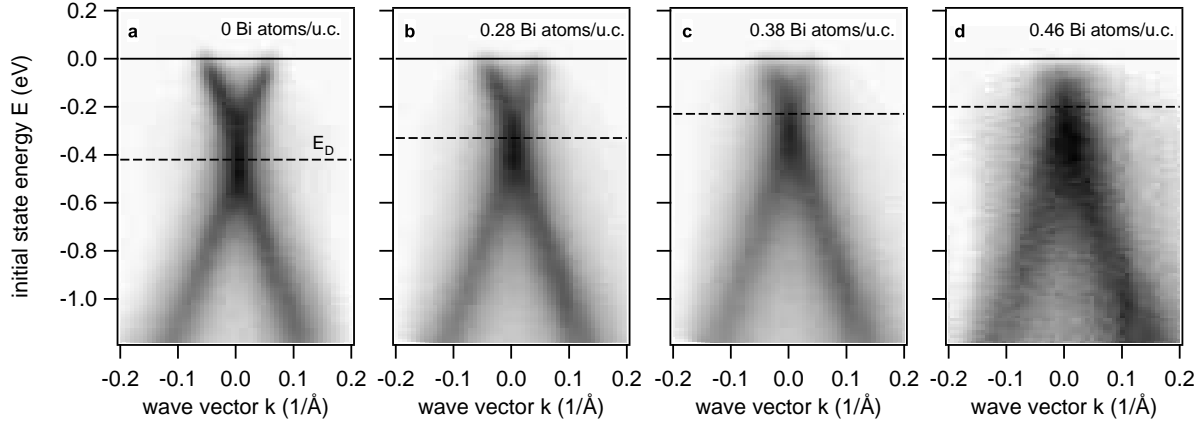}
  \caption{{\bf Doping graphene with Bi atoms:} experimental band structure of epitaxial graphene doped with
  bismuth atoms. (a) pristine graphene layer, (b)-(d) increasing
  amounts of bismuth atoms have been deposited.}
  \label{fig:Data}
\end{figure*}

We have studied the valence band structure of a single graphene
layer on 4H-SiC(0001) using angle-resolved photoemission
spectroscopy. Fig.\ \ref{fig:Data} shows the experimental band
structure of epitaxial graphene doped with successively higher
amounts of Bi atoms. The initial state energy $E$ of the bands is
plotted as a function of the electron wave vector $k$ \cite{GK}.
The intensity scale is linear with light and dark areas
corresponding to low and high photoelectron current, respectively.
The Dirac point is located at the $\overline{\mbox{K}}$-point,
which lies in the corner of the hexagonal Brillouin zone. Fig.\
\ref{fig:Data}(a) shows the pristine graphene layer. The linear
dispersion of the valence and conduction bands is clearly visible.
Due to the charge transfer with the SiC-substrate, the Dirac cone
of the conduction band is partially filled shifting the Dirac
point into the occupied states by about 420\,meV
\cite{Bostwick,Zhou,Riedl}. In Figs.\ \ref{fig:Data} (b)--(d)
successively higher amounts of bismuth atoms per graphene unit
cell (u.c.) as indicated in each panel are deposited on the
graphene layer \cite{graphene}. As the bismuth coverage increases
the Dirac point clearly shifts towards the Fermi level. Otherwise,
the band structure remains unaltered by the bismuth adatoms, i.~e.
the linear dispersion is preserved. Only at high bismuth coverage
the line width of the bands increases noticeably. This is probably
related to the Bi atoms not forming an ordered structure on the
surface, which leads to broadening of the photoemission features.
Furthermore, the number of free charge carriers decreases as a
successively smaller cross section of the conduction band
intersects the Fermi level.

Very similar results have been obtained for antimony atoms
deposited on the graphene layer. Antimony is also located in group V of the periodic table just above bismuth, so that a very
similar doping behavior is expected. The experimental band
structure is not shown here, but looks very much like the data
obtained for bismuth on graphene except that it takes a higher
antimony coverage to reach the same doping level.

\begin{figure}
  \includegraphics[width = 0.5\textwidth]{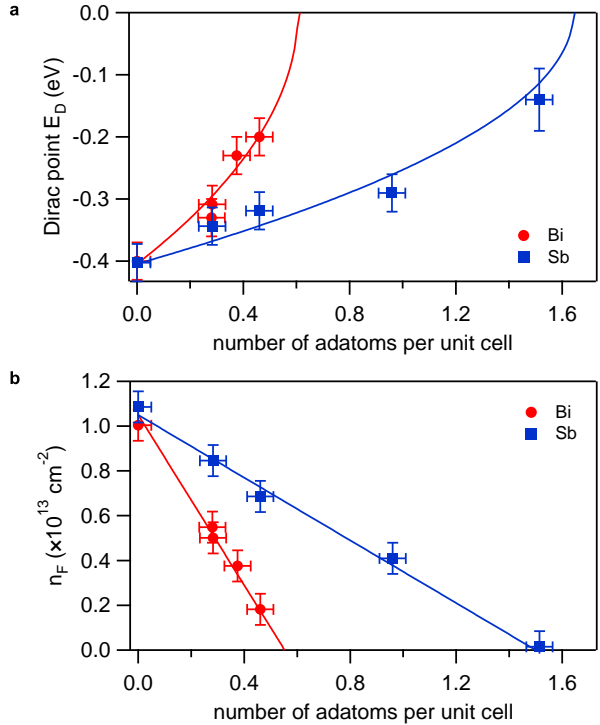}
  \caption{{\bf Doping parameters of Bi and Sb:} (a) position of the Dirac point $E_D$ and (b) free charge
  carrier density $n_F$ as a function of doping with bismuth atoms (red circles)
  and antimony atoms (blue squares) (adatoms per graphene unit cell). The solid lines represent a simple model calculation
  assuming an electron transfer of 0.01 and 0.0036 electrons per Bi and Sb atom, respectively.}
  \label{fig:Analysis}
\end{figure}

A more quantitative analysis of the bismuth and antimony doping is
displayed in Fig.\ \ref{fig:Analysis}. Panel (a) shows the evolution of the Dirac point as
a function of the coverage, which is given in number of atoms
per graphene unit cell. The Dirac point clearly approaches the
Fermi level with increasing doping indicating that there is charge
transfer from the graphene layer to the adatoms. A simple
theoretical model based on the linear density of states for the
graphene layer has been used to estimate the doping effect of the
bismuth atoms assuming that the charge transfer is proportional to
the amount of dopant atoms:
\begin{equation}
E_D=-\sqrt{\pi}\hbar v_F\sqrt{N_0-N_h}
\end{equation}
Here $E_D$ is the Dirac point, with the zero of the energy scale
referenced to the Fermi level. The Fermi velocity is $\hbar
v_F=6.726$\,eV\AA\ \cite{Geim,Bostwick}, $N_0$ is the number of
electrons in the conduction band for zero doping, and $N_h$ is the
number of holes doped into the graphene layer. If we assume that
about 0.01 electrons per bismuth atom and 0.0036 electrons per
antimony atom are extracted from the graphene layer, the Dirac
point should follow the respective solid lines plotted in Fig.\
\ref{fig:Analysis}(a). It would reach the Fermi level for a
coverage of 0.61 bismuth atoms and 1.65 antimony atoms per unit
cell. The experiment shows, however, that for higher coverages the
bands become broader and less well-defined. At these points the
average distance between adatoms approaches the one of monolayer coverage, so
that the photoelectrons will interact and scatter in the adlayer.

\begin{figure}
  \includegraphics[width = 0.3\textwidth]{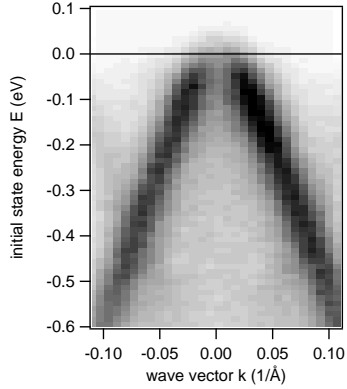}
  \caption{{\bf Hole doping with Au:} experimental band structure of Au atoms on epitaxial graphene. The bands
  are well defined with the Dirac point at about 100\,meV above the Fermi level and a
  charge carrier density for the holes of about $5\times10^{11}\,$cm$^{-2}$.}
  \label{fig:AuDoping}
\end{figure}

In Fig.\ \ref{fig:Analysis}(b), the free charge carrier density of
the graphene layer is plotted as a function of adatoms per
unit cell. It has been extracted from the experimental data
through the Fermi wave vector $k_F$ via $n_F=k_F^2/\pi$, a formula which relates Fermi wave vector $k_F$ and free charge carrier density $n_F$ for two-dimensional electron gases. The charge carrier density is
clearly reduced as the number of adatoms increases indicating hole
doping. The solid line shows a linear dependence of the charge
carrier density as a function of bismuth coverage with a very good
correspondence to the experimental data assuming the same values
for the electron transfer per adatom as above.

For actual p-type doping with holes as charge carriers, the Dirac
point has to shift into the unoccupied states. A further increase
of the electron transfer from the graphene layer to the adatoms
is desirable. The natural starting point would be an element with
a higher electron affinity than bismuth or antimony. Motivated by
this, we have deposited gold atoms on epitaxial graphene, as its
electron affinity is about twice as high as for bismuth and as recent phototransport experiments indicated that gold contacts induce p-doping in graphene \cite{Lee}. The experimental band structure for about two gold
atoms per graphene unit cell shown in (Fig.\ \ref{fig:AuDoping}) clearly displays p-type doping. Both branches of the valence band cone
clearly cross the Fermi level close to the
$\overline{K}$-point leaving the valence band partially
unoccupied. We estimate the Dirac point to be about 100\,meV above
the Fermi level and a free charge carrier density of the holes of
about $5\times 10^{11}$\,cm$^{-2}$.

The bands in Fig.\ \ref{fig:AuDoping} are much narrower than for
the bismuth adatoms in Fig.\ \ref{fig:Data}(d) even though the
gold coverage is much higher than the bismuth coverage. The sharp bandstructure suggests that gold forms an ordered structure on the graphene
layer. Furthermore, the p-type doping in Fig.\ \ref{fig:AuDoping}
is only induced after a post-annealing of the sample to at least
700$^{\circ}$C. Similar to what has been observed for gold
atoms binding to pentacene \cite{Repp}, an actual bond between the gold
atom and the graphene is formed after a certain activation barrier
is overcome. The chemical bond formation is confirmed by x-ray photoelectron spectroscopy results, showing a clear splitting of the carbon 1s core level after annealing the sample at 700$^{\circ}$C, which is not present for the clean graphene layer. Interestingly, while this goes beyond the idea of
surface transfer doping, where a covalent bond between adatom and
graphene layer is not present, the peculiar band structure of
graphene with its linear dispersion remains intact. This clearly
demonstrates that for a coverage of about two gold atoms per
graphene unit cell the Dirac point can be shifted above the Fermi
level leaving the valence band partially unoccupied. Surprisingly, the estimated electron transfer is only 0.0024 electrons per gold atom, which is only a quarter of the value found for bismuth, where, however, no chemical bond to the graphene layer is formed.

Our results demonstrate that p-type doping of an epitaxial
graphene layer is possible by means of simple atoms. While bismuth
and antimony were only able to shift the Dirac point in the
direction of the Fermi level, i.~e. reduce the natural n-type
doping of the substrate, gold actually shifted the Dirac point
into the unoccupied states thereby inducing p-type doping.
Epitaxial graphene on silicon carbide becomes a feasible
alternative to conventional electronic materials as n-type doping
is naturally induced and p-type doping can be achieved by doping
with gold atoms, which are easily processed. With its potential for large scale production \cite{Kedzierski} all the advantages of epitaxial graphene, e.~g.
the strict two-dimensionality, high carrier mobility, high current
densities and ballistic transport at room temperature \cite{Geim},
are available for device application.

\section*{Methods}
Photoemission experiments have been done in ultra-high-vacuum (UHV). The UHV system is equipped with a hemispherical SPECS HSA3500 electron analyser with an energy resolution of $\sim$10meV. We used HeII radiation with an energy of 40.8eV for the ARUPS measurements. Our n-doped 4H-SiC(0001) substrate was prepared by hydrogen-etching and subsequent Si deposition at 800$^{\circ}$C until a homogeneous and sharp ($3\times 3$) LEED pattern appeared. A monolayer graphene was obtained after annealing for 5min at 1150$^{\circ}$C. Bi and Sb were deposited on a room temperature sample using a commercial electron beam evaporator which was calibrated with the help of the ($\sqrt3\times\sqrt3$) reconstruction that is formed for 1/3 monolayer coverage of both Bi and Sb on Ag(111). Au was deposited at room temperature with a commercial Knudsen cell which was calibrated using a quartz crystal microbalance. After Au deposition the sample was annealed at 700$^{\circ}$C for 5min. All measurements were conducted at room temperature.

%
% Back matter
%


\begin{thebibliography}{12}
\bibitem{GK} The measurements were
taken at the $\overline{\mbox{K}}$-point with the wave vector axis
perpendicular to the $\overline{\Gamma}\overline{\mbox{K}}$-high
symmetry line.
\bibitem{graphene} The graphene unit cell used here has a lattice constant of 2.46\,\AA\ (hexagonal lattice) and contains two graphene atoms.
\end{thebibliography}
\end{document}